\documentclass[sigconf,9pt]{acmart}

\usepackage{graphicx}
\usepackage{url}
\usepackage{color}
\usepackage{alltt}
\usepackage{syntax}
\usepackage{newfloat}
\usepackage{mdframed}
\usepackage{multicol}
\usepackage{amsmath}
\usepackage{tikz}
\usepackage{filecontents} 
\usepackage{graphicx}
\usepackage{epstopdf} 
\usepackage{xspace}
\usepackage{enumitem}
\usepackage{svg}
\usepackage{lscape}
\usepackage{longtable}
\usepackage{supertabular}

\usepackage[ruled,linesnumbered,noend]{algorithm2e}
\SetKw{Continue}{continue}
\SetKwInput{KwInput}{Input}
\SetKwInput{KwOutput}{Output}
\SetKwProg{Fn}{Function}{}{end}

\SetCommentSty{mycommfont}

\usepackage{adjustbox}
\usepackage{wrapfig}
\usepackage{graphbox}
\usepackage{rotating}
\usepackage{pdfpages}
\usepackage{mdwlist}
\usepackage{multirow}
\usepackage[T1]{fontenc}

\usepackage{subcaption}
\usepackage[labelfont=bf,skip=0pt]{caption}
\usepackage{hyperref} 
\hypersetup{
	citebordercolor={0 1 0},
	citecolor=blue,
	colorlinks=false,
	filebordercolor={0 .5 .5},
	filecolor=green,
	linkbordercolor={1 0 0},
	linkcolor=blue,
	menubordercolor={1 0 0},
	pageanchor=true,
	pagebackref=true,
	pdfborder={0 0 1},
	pdfpagelabels=true,
	pdftex,
	plainpages=false,
	urlbordercolor={0 1 1},
	urlcolor=blue,
}

\usepackage[capitalize,nameinlink]{cleveref}
\hypersetup{%
	bookmarksnumbered, bookmarksopen=true, bookmarksopenlevel=1,%
}
\crefname{figure}{Figure}{Figures}
\crefname{listing}{Query}{Queries}
\crefname{section}{Section}{Sections}
\crefname{table}{Table}{Tables}
\crefname{BNF}{Grammar}{Grammars}
\crefname{algorithm}{Algorithm}{Algorithms}
\crefname{equation}{Equation}{Equations}

\usepackage{listings}
\definecolor{mygreen}{rgb}{0,0.6,0}
\definecolor{mygray}{rgb}{0.5,0.5,0.5}

\newcommand{\distance}{5pt}
\DeclareFloatingEnvironment[
fileext   = logr,
listname  = {List of Grammars},
name      = Grammar,
placement = !htbp
]{BNF}

\newcommand{\msim}{\raise.17ex\hbox{$\scriptstyle\sim$}}

\newcommand{\myparatight}[1]{\smallskip\noindent{\bf {#1}.}}

\newcommand{\eat}[1]{}
\newcommand{\eg}{e.g.,\xspace}
\newcommand{\ie}{i.e.,\xspace}

\newcommand{\tool}{\textsc{SecurityKG}\xspace}
\newcommand{\cti}{OSCTI\xspace}
\newcommand{\seckg}{security knowledge graph\xspace}

\AtBeginDocument{%
  \providecommand\BibTeX{{%
    \normalfont B\kern-0.5em{\scshape i\kern-0.25em b}\kern-0.8em\TeX}}}

\setcopyright{acmcopyright}
\copyrightyear{2021} 
\acmYear{2021} 
\setcopyright{acmlicensed}\acmConference[SIGMOD '21]{Proceedings of the 2021 International Conference on Management of Data}{June 18--27, 2021}{Virtual Event , China}
\acmBooktitle{Proceedings of the 2021 International Conference on Management of Data (SIGMOD '21), June 18--27, 2021, Virtual Event , China}
\acmPrice{15.00}
\acmDOI{10.1145/3448016.3452745}
\acmISBN{978-1-4503-8343-1/21/06}

\begin{document}

\fancyhead{}

\title{A System for Automated Open-Source Threat Intelligence Gathering and Management}

\author{%
Peng Gao$^{1*}$, Xiaoyuan Liu$^{1*}$, Edward Choi$^1$, Bhavna Soman$^2$, Chinmaya Mishra$^2$, Kate Farris$^2$, Dawn Song$^1$
}

\affiliation{%
    \normalsize{
        \institution{$^*$Equal Contribution\\ $^1$University of California, Berkeley\; $^2$Microsoft Corporation}
        \country{\{penggao,xiaoyuanliu,edwardc1028,dawnsong\}@berkeley.edu, \{Bhavna.Soman,Chinmaya.Mishra,Kate.Farris\}@microsoft.com}
    }
}

\renewcommand{\shortauthors}{Gao, et al.}

\begin{abstract}

To remain aware of the fast-evolving cyber threat landscape, open-source Cyber Threat Intelligence (\cti) has received growing attention from the community.
Commonly, knowledge about threats is presented in a vast number of \cti reports.
Despite the pressing need for high-quality \cti, existing \cti gathering and management platforms, however, have primarily focused on isolated, low-level Indicators of Compromise.  
On the other hand, higher-level concepts (\eg adversary tactics, techniques, and procedures) and their relationships have been overlooked, which contain essential knowledge about threat behaviors that is critical to uncovering the complete threat scenario.
To bridge the gap, we propose \tool, a system for automated \cti gathering and management.
\tool collects 
\cti reports from various sources, uses a combination of AI and NLP techniques to extract high-fidelity knowledge about threat behaviors, and constructs a 
\seckg.
\tool also provides a UI that supports various types of interactivity to facilitate knowledge graph exploration.

\end{abstract}



\maketitle

\section{Introduction}
\label{sec:intro}

\begin{figure*}[t]
    \centering
    \includegraphics[width=\linewidth]{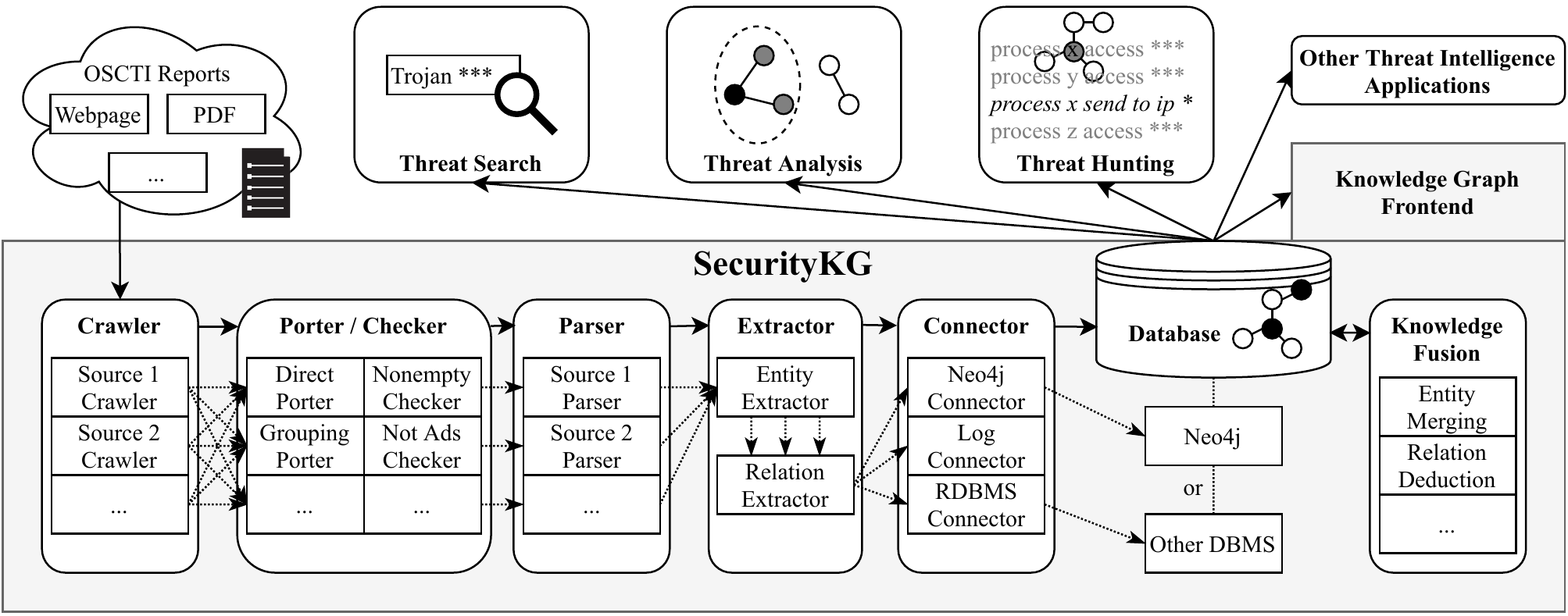}
    \caption{The architecture of \tool. Arrows represent data flows.}
    \label{fig:architecture}
\end{figure*}

Sophisticated cyber attacks have plagued many high-profile businesses~\cite{equifax}.
To remain aware of the fast-evolving threat landscape and gain insights into the most dangerous threats, open-source Cyber Threat Intelligence (\cti)~\cite{li2019reading} has received growing attention from the community.
Commonly, knowledge about threats 
is presented in a vast number of \cti reports in various forms (\eg threat reports, security news and articles~\cite{securelist,phishtank}).
Despite the pressing need for high-quality \cti,
existing \cti gathering and management systems~\cite{threatminer,threatcrowd,alienvault-otx}, 
however, have primarily focused on simple Indicators of Compromise (IOCs)~\cite{liao2016acing}, such as signatures of artifacts, malicious file/process names, IP addresses, and domain names.
Though effective in capturing isolated, low-level IOCs, these platforms cannot capture higher-level behaviors such as adversary tactics, techniques, and procedures
\cite{mitre-attack}, which are tied to the attacker’s goals and thus much harder to change.
As the volume of OSCTI sources increases day-by-day, it becomes increasingly challenging to maneuver through and correlate the myriad of sources to gain useful insights.
Towards this end, there is a pressing need for a new system that can harvest and manage high-fidelity threat intelligence in an automated, intelligent, and principled way.

There are several major challenges for building such a system.
First, \cti reports come in diverse formats: some reports contain structured fields such as tables and lists, and some reports primarily consist of unstructured natural-language texts.
The platform is expected to be capable of handling such diversity and extracting information.
Second, besides IOCs, \cti reports contain various other entities that capture threat behaviors.
The platform is expected to have a wide coverage of entity and relation types to comprehensively model the threats.
Third, accurately extracting threat knowledge from unstructured \cti texts is non-trivial.
This is due to the presence of massive nuances particular to the security context, such as special characters (\eg dots, underscores) in IOCs.
These nuances limit the performance of most NLP modules (\eg sentence segmentation, tokenization).
Besides, some learning-based information extraction approaches require large annotated training corpora,
which is expensive to obtain manually. 
Thus, how to 
programmatically 
obtain annotations becomes another challenge.

To bridge the gap, we built \tool ($\sim9$K lines of Python code), a system for automated \cti gathering and management. 
\tool
collects \cti reports from various sources, uses a combination of AI and NLP techniques to extract high-fidelity knowledge about threat behaviors as security-related entities and relations, constructs a \emph{security knowledge graph} containing the entity-relation triplets, and updates the knowledge graph by continuously ingesting new data.
Specifically, \tool has the following key components:
(1) a set of fast and robust crawlers for collecting OSCTI reports from $40$+ major security websites;
(2) a 
security knowledge ontology that models a wide range of high-level and low-level security-related entities (\eg IOCs, malware, threat actors, techniques, tools) and relations; 
(3) a combination of AI and NLP techniques (\eg Conditional Random Fields~\cite{lafferty2001conditional}) 
to accurately extract entities and relations; specifically, we leverage data programming~\cite{ratner2016data} to programatically create large training corpora;
(4) an
extensible backend system that manages all components for OSCTI gathering, knowledge extraction, and knowledge graph construction and persistence;
(5) a 
UI that provides various types of interactivity to facilitate knowledge graph exploration.

Different from general knowledge graphs~\cite{miller1995wordnet,auer2007dbpedia,mahdisoltani2013yago3} that store and represent general knowledge (\eg movies, actors), \tool targets automated extraction and management of \cti knowledge for the security domain. 
\tool is the first work in this space.

\textbf{Demo video:} \url{https://youtu.be/8PDJSaTnLDc}

\section{\tool Architecture}
\label{sec:design}

\cref{fig:architecture} shows the architecture of \tool.
\tool manages the 
lifecycle of security knowledge in four stages: collection (Crawler), processing (Porter/Checker, Parser, Extractor), storage (Connector, Database), and applications.
In the collection stage, \tool periodically and incrementally collects \cti reports from multiple sources.
In the processing stage, \tool parses the reports, extracts structured knowledge, and constructs a \seckg based on a pre-defined 
ontology.
In the storage stage, \tool inserts the knowledge into 
backend databases for storage.
Various applications (\eg threat searching, threat analysis, threat hunting) can be built by accessing the \seckg stored in the databases.
\tool also provides a frontend UI to facilitate knowledge graph exploration.

\subsection{Backend System Design}
\label{subsec:backend}

To handle 
diverse \cti reports, the system needs to be scalable, and 
maintain a unified representation of all possible knowledge types in both known and future data sources.
The system also needs to be 
extensible to incorporate new data sources and processing and storage components to serve the needs of different applications.

\myparatight{Scalability}
To make the system scalable, we parallelize the processing procedure of \cti reports.
We further pipeline the processing steps in the procedure to improve the throughput.
Between different 
steps in the pipeline, we specify the formats of intermediate representations and make them serializable.
With such pipeline design, we can have multiple computing instances for a single step and pass serialized intermediate results across the network, making multi-host deployment and load balancing possible.

\myparatight{Unified Knowledge Representation}
To comprehensively represent security knowledge, we design an \emph{intermediate CTI representation} and separate it from the security knowledge ontology.
Intermediate CTI representation is a schema that covers relevant and potentially useful information in all data sources and lists out corresponding fields.
We construct this schema by iterating through data sources, adding previously undefined types of knowledge, and merging similar fields. 
Specifically, our \emph{source-dependent parsers} will first convert the original \cti reports into representations (\ie Python objects in memory) that follow this schema by parsing the structured fields (\eg fields identified by HTML tags).
Then, our \emph{source-independent extractors} will further refine the representations by extracting information (\eg IOCs, malware names) from unstructured texts and putting it into the corresponding fields. 

Directly using 
these intermediate representations results in inefficient storage.
Furthermore, these long representations are not convenient for end users (\eg threat analysts) to analyze.
Thus, before merging 
them into the storage through connectors, \tool refactors them to match our security knowledge ontology, which is separately designed and 
has clear and concise semantics.

\myparatight{Extensibility}
To make the system extensible, we adopt a modular design, allowing multiple components with the same interface to work together in the same processing step.
For example, \tool by default uses a Neo4 connector to export knowledge into a Neo4j database~\cite{neo4j}.
However, if the user cares less about multi-hop relations, he/she may switch to a RDBMS using a SQL connector.
Similarly, parsers and extractors can be switched or extended (sharing the same input/output formats), 
making the system extendable. 
Furthermore, the system can be configured through a user-provided configuration file, which specifies the set of components to use and the additional parameters (\eg threshold values for entity recognition) that are passed to these components.

\subsection{\cti Reports Collection}
\label{subsec:collection}

We built a crawler framework that has $40$+ crawlers for collecting \cti reports from major security sources (each crawler handles one data source), covering threat encyclopedias, blogs,
security news, etc.
The crawler framework 
schedules periodic execution and reboot after failure for different crawlers in an efficient and robust manner.
It also has a multi-threaded design to boost the efficiency, 
achieving a throughput of approximately $350$+ reports per minute on a single deployed host.
In total, we have collected over $120$K+ \cti reports and the number is still increasing.

\subsection{Security Knowledge Ontology Design}
\label{subsec:ontology}

\begin{figure}[t]
    \centering
    \includegraphics[width=\linewidth]{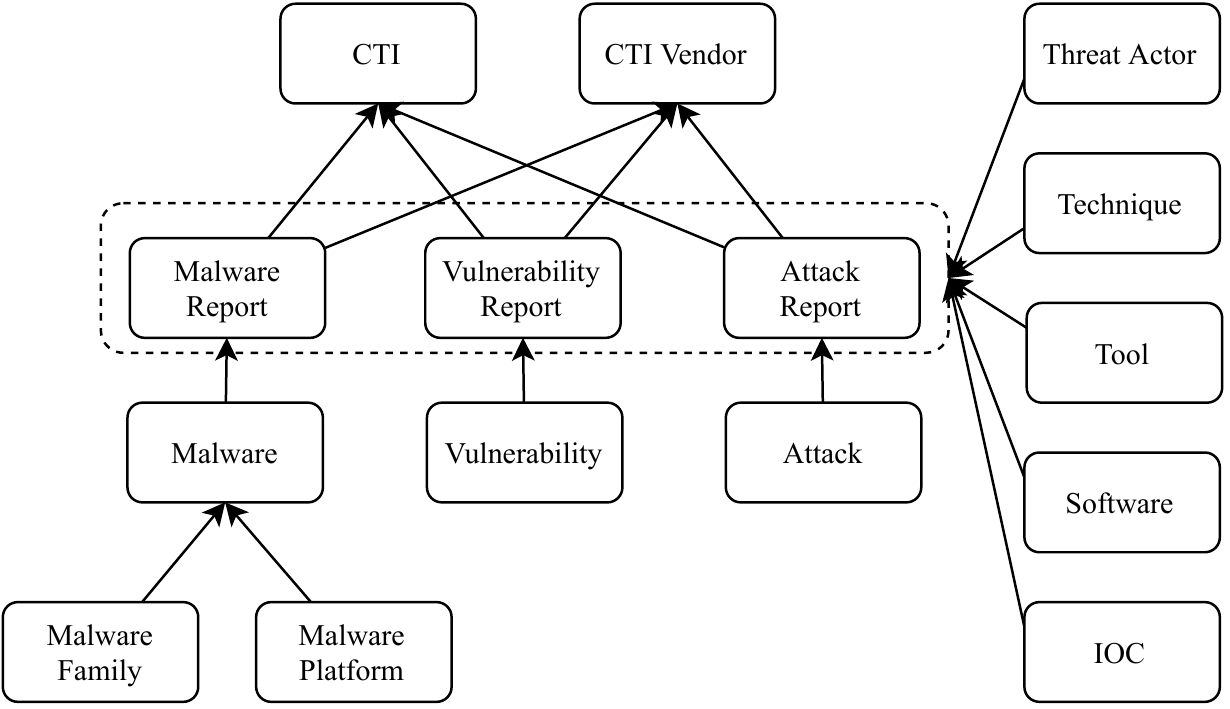}
    \caption{Security knowledge ontology}
    \label{fig:ontology}
\end{figure}

\cref{fig:ontology} shows our security knowledge ontology, which specifies the types of security-related entities and relations in the \seckg.
Based on our observations of \cti data sources, we categorize \cti reports into three types: malware reports, vulnerability reports, and attack reports.
For each report, we associate it with an entity of the corresponding type.
Besides, reports are created by specific CTI vendors, and often contain information concepts on threat actors, techniques, tools, software, and various types of IOCs (\eg file name, file path, IP, URL, email, domain, registry, hashes). 
Thus, we create entities for these concepts as well.
Entities have relationships between them (\eg <MALWARE\_A, DROP, FILE\_A> specifies a ``DROP'' relation between a ``MALWARE'' entity and a ``FILE'' entity), as well as attributes in the form of key-value pairs.
By constructing such an ontology, we can capture different types of security knowledge in the system.
Compared to other cyber ontologies~\cite{stix,syed2016uco}, our ontology targets a larger set. 
\cref{fig:ui} shows an example knowledge subgraph 
that follows this ontology.

\subsection{Security Knowledge Extraction}
\label{subsec:extraction}

We describe the steps inside the processing stage that extract security knowledge from the collected \cti report files (\eg HTML, PDF).
The porters take the input report files and convert them into \emph{intermediate report representations}; they group multi-page reports and add metadata like ids, sources, titles, and original file locations and timestamps.
The checkers work as filters on the list of intermediate report representations; they screen out irrelevant reports like empty pages or ads by running condition checks.
The parsers are source-dependent, taking the advantage of prior knowledge of the source website structure and extracting keys and values from report files. 
They convert the list of intermediate report representations into a list of \emph{intermediate CTI representations} (\cref{subsec:backend}).
The extractors further refine these intermediate CTI representations by completing some of the 
fields using entity recognition and relation extraction.
Since the intermediate CTI representation is a unified format, the extractors are source-independent.

Next, we describe the design of the extractors.

\myparatight{Security-Related Entity Recognition}
We adopt a Conditional Random Field (CRF)~\cite{lafferty2001conditional} model to extract security-related entities in unstructured texts.
Compared to general named entity recognition, we are faced with two unique challenges: (1) presence of massive nuances particular to the security context; (2) lack of large annotated training corpora.
To address the first challenge, as these nuances mostly exist in IOCs, we use a method called \emph{IOC protection} proposed in our other work~\cite{gao2021enabling}, 
by replacing IOCs with meaningful words in natural language context 
(\eg the word ``something'') and restoring them after the tokenization procedure.
This way, we guarantee that the potential entities are complete tokens.
To address the second challenge, we programmatically synthesize annotations using data programming~\cite{ratner2016data}.
Particularly, we create labeling functions based on our curated lists of entity names. 
For example, the lists of threat actors, techniques, and tools are constructed from MITRE ATT\&CK~\cite{mitre-attack}.
To train the CRF model, we use features such as word lemmas, pos tags, and word embeddings~\cite{mikolov2013distributed}. 
Since our model has the ability to leverage token-level semantics,
it can outperform a naive entity recognition solution that relies on regex rules, and generalize to entities that are not in the training set.

\myparatight{Security-Related Relation Extraction}
To extract relations between security-related entities, 
since it is relatively hard to programmatically synthesize annotations for relations,
we adopt an unsupervised approach.
In particular, we leverage the dependency-parsing-based IOC relation extraction pipeline proposed in our other work~\cite{gao2021enabling}, and extend it to support the extraction of relation verbs between entities recognized by our CRF model.

\subsection{Security Knowledge Graph Construction}
\label{subsec:graph}

As a final step, \tool inserts the processed results into the backend storage using connectors. 
The connector 
merges the intermediate CTI representations into the corresponding storage backend by refactoring them to match our security knowledge ontology, such that the previously constructed security knowledge graph can be augmented with new knowledge.

Since we store the knowledge extracted from a large number of reports in the same knowledge graph, one potential problem is that nodes constructed from different reports may refer to the same entity.
We made the design choice that, in this step, we only merge nodes with exactly the same description text. 
It is possible that nodes with similar description texts actually refer to the same entity (\eg same malware represented in different naming conventions by different CTI vendors).
For these nodes, we merge them in a separate knowledge fusion stage, by creating a new node with unified attributes and migrating all relation edges.
By separating the knowledge fusion stage from the storage stage in the main pipeline, we can prevent early deletion of useful information.


\subsection{Frontend UI Design}
\label{subsec:ui}

\begin{figure}[t]
    \centering
    \includegraphics[width=\linewidth]{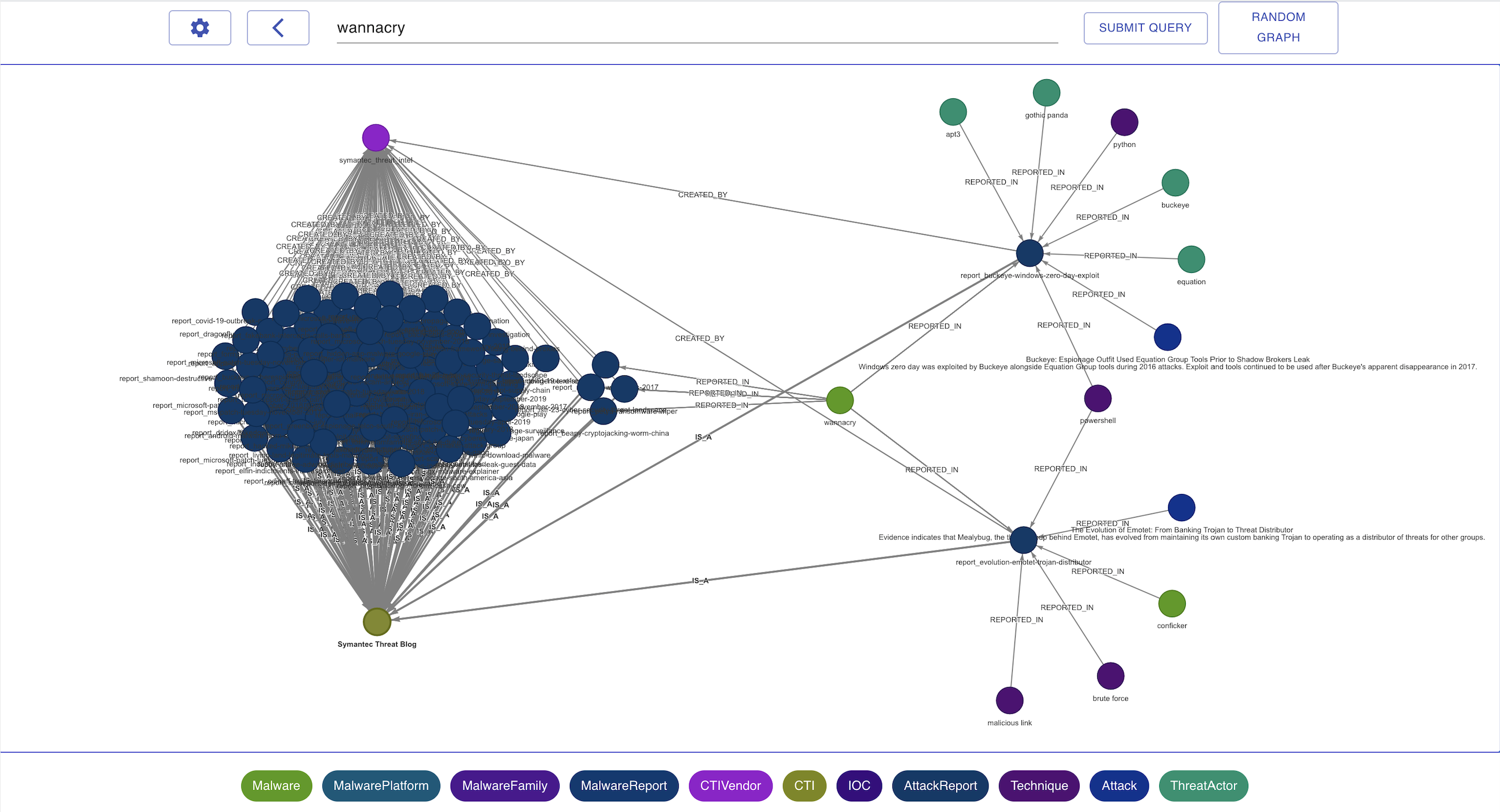}
    \caption{The web UI of \tool}
    \label{fig:ui}
\end{figure}

In order to facilitate knowledge graph exploration,
we built a web UI using React and Elasticsearch.
\cref{fig:ui} shows an example subgraph of security knowledge graph in our UI. 
Currently, the UI interacts with the Neo4j database through a Neo4j JS driver, 
and provides various functionalities to facilitate the exploration of the knowledge graph, which we  describe next.

We built features to simplify user view. 
The user can zoom in and out and drag the canvas.
Node names and edge types are 
displayed by default.
Nodes are colored according to their types.
When a node is hovered over, its detailed information will be displayed.

We built features that facilitate threat search and knowledge graph exploration. 
First, the UI provides multilingual query support so that the user can search information using keywords (through Elasticsearch) or Cypher queries (through Neo4j Cypher engine), which enables the user to easily identify targeted threats in the large graph.
Second, the user can drag nodes around on the canvas.
The UI actively responds to node movements to prevent overlap through an automatic graph layout using the Barnes-Hut algorithm~\cite{barnes1986hierarchical}, which calculates the nodes' approximated repulsive force based on their distribution. 
The dragged nodes will lock in place but are still draggable if selected.
This node draggability feature helps the user define custom graph layouts.
Third, the UI supports inter-graph navigation.
When a node is double-clicked, if its neighboring nodes have not appeared in the view yet, these neighboring nodes will automatically spawn.
On the contrary, once the user is done investigating a node, if its neighboring nodes or any downstream nodes are shown, double clicking on the node again will hide all its neighboring nodes and downstream nodes. 
This node expansion/collapse feature is essential for convenient graph exploration.

We built features that provide flexibility to the user.
The user can configure the number of nodes 
displayed
and the maximum number of neighboring nodes displayed for a node.
The user can view the previous graphs displayed
by clicking on the back button.
The user can also fetch a random subgraph for exploration.

\section{Demonstration Outline}
\label{sec:demo}

In our demo, we first show various usage scenarios of \tool's UI.
Specifically, we perform two keyword searches and one Cypher query search and demonstrate all supported features:

\begin{itemize}[listparindent=\parindent, leftmargin=*]
    \item \emph{Keyword search for ``wannacry''}: 
    We first investigate the WannaCry ransomware by performing a keyword search.
    Throughout the investigation, we aim to demonstrate functionalities including detailed information display, node dragging, automatic graph layout, canvas zooming in/out, and node expansion/collapse.
    We will end the investigation with a subgraph that shows all the relevant information (entities) of the WannaCry ransomware.

    \item \emph{Keyword search for ``cozyduke''}: 
    In the second scenario, we perform a keyword search of a threat actor, CozyDuke.
    We will investigate the techniques used by CozyDuke, and check if there are other threat actors that use the same set of techniques.

    \item \emph{Cypher query search}: 
    In the third scenario, we execute a specific Cypher query, {\tt match(n) where n.name = ``wannacry'' return n}, to demonstrate that the same WannaCry node will be returned as in the first scenario.
    We then execute other queries.
    
\end{itemize}

Our demo video gives a 
walkthrough of these scenarios.
In addition to threat search and knowledge graph exploration, we demonstrate the end-to-end automated data gathering and management procedure of \tool. 
We will empty the database and apply \tool to a number of \cti sources. 
We will demonstrate various system components, and provide insights into how \cti reports are collected, how entities and relations are extracted, and how information is merged into the knowledge graph so that the graph can continuously grow.
The audience will have the option to try the UI and the whole system to gain deeper insights into various system components and the supported functionalities.

\section{Related Work}

Besides existing \cti gathering and management systems~\cite{threatminer,threatcrowd,alienvault-otx}, research progress has been made to better analyze \cti reports, including extracting IOCs~\cite{liao2016acing}, extracting threat action terms from semi-structured Symantec reports~\cite{husari2017ttpdrill}, understanding vulnerability reproducibility~\cite{mu2018understanding}, and measuring threat intelligence quality~\cite{li2019reading,dong2019towards}. 
Research has also proposed to leverage individual \cti reports for threat hunting~\cite{gao2021enabling}. 
\tool distinguishes from all these works in the sense that it targets automated construction of a knowledge graph particularly for the security domain, by extracting a wide range of security-related entities and relations from a large number of \cti reports using AI and NLP techniques.

In future work, we plan to connect \tool with our query-based threat protection systems (\eg attack investigation~\cite{gao2018aiql,gao2019query}, attack detection~\cite{gao2018saql,gao2020querying}, threat hunting~\cite{gao2021enabling,gao2021system}) to enable knowledge-enhanced cyber threat protection.

\section{Conclusion}

We have presented \tool, a system for automated \cti gathering and management.
\tool uses a combination of AI and NLP techniques to extract threat knowledge from a large number of collected \cti reports, and constructs a security knowledge graph to structuralize and persist the knowledge.
\tool has potential to empower a variety of security applications.

\myparatight{Acknowledgement}
This work was supported by the 2020 Microsoft Security AI RFP Award, the Azure cloud computing platform, and the UC Berkeley Center for Long-Term Cybersecurity (CLTC).

\bibliographystyle{ACM-Reference-Format}
\bibliography{refs}

\end{document}